\documentclass{aip-cp}

\usepackage[numbers,compress]{natbib}
\usepackage{rotating}
\usepackage{graphicx}
\usepackage{siunitx}

\begin{document}

\title{Positron Beams and Two-Photon Exchange: The Key to Precision Form Factors}

\author[mit]{Jan C.\ Bernauer}\corref{cor1}

\affil[mit]{Massachusetts Institute of Technology, 77 Massachusetts Avenue, Cambridge, MA 02139, USA}
\corresp[cor1]{Corresponding author: bernauer@mit.edu}

\maketitle

\begin{abstract}
The proton elastic form factor ratio can be measured either via Rosenbluth separation in an unpolarized beam and target experiment, or via the use of polarization degrees of freedom. However, data produced by these two approaches show a discrepancy, increasing with $Q^2$. The proposed explanation of this discrepancy---two-photon exchange---has been tested recently by three experiments. The results support the existence of a small two-photon exchange effect but cannot establish that theoretical treatment at the measured momentum transfers are valid. At larger momentum transfers, theory remains untested. This paper investigates the possibilities of measurements at DESY and Jefferson Lab to measure the effect at larger momentum transfers.
\end{abstract}

\section{INTRODUCTION}
\begin{figure}[htb]
  \centerline{\includegraphics[width=\textwidth]{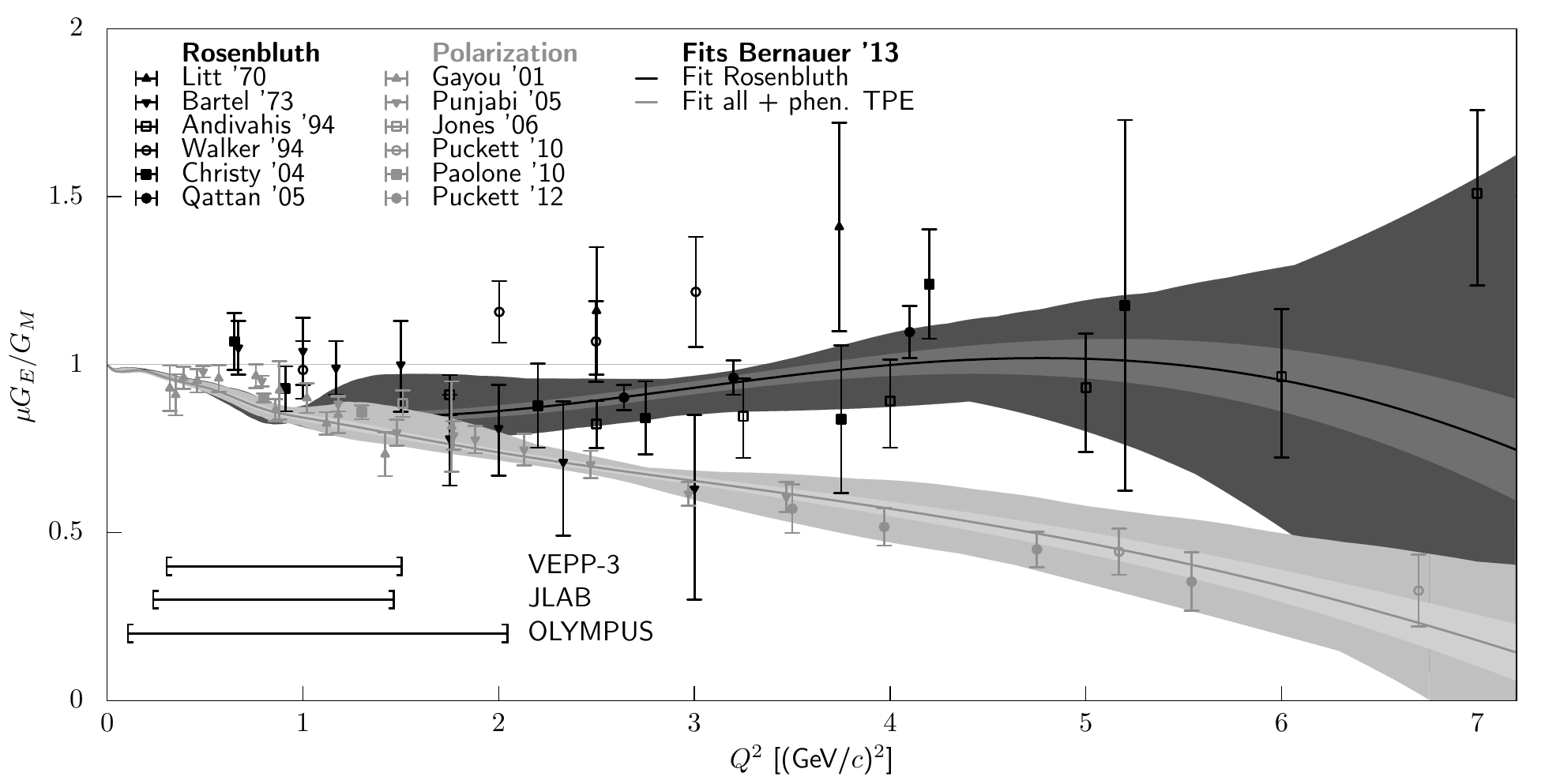}}
  \caption{\label{figratio}The proton form factor ratio $\mu G_E/G_M$, as determined via Rosenbluth-type (black points, from \cite{litt,bartel,andivahis,walker,christy,qattan}) and polarization-type (gray points, from \cite{gayou,punjabi,jones,puckett10,paolone,puckett12}) experiments. While the former indicate a ratio close to 1, the latter show a distinct linear fall-off. Curves are from a phenomenological fit \cite{bernauer13}, to either the Rosenbluth-type world data set alone (dark curves) or to all data, then including a phenomenological TPE model.}
\end{figure}

Over more than half a century, proton elastic form factors have been studied in electron-proton scattering with unpolarized beams. These experiments have yielded data over a large range of four-momentum transfer squared, $Q^2$. The form factors were extracted from the cross sections via the so-called Rosenbluth separation. Among other things, they found that the form factor ratio $\mu G_E/G_M$ is in agreement with scaling, i.e., that the ratio is constant. Somewhat more recently, the ratio of the form factors was measured using polarized beams, with different systematics and increased precision especially at large $Q^2$. However, the results indicate a roughly linearly fall-off of the ratio. This discrepancy can be seen in Fig.\ \ref{figratio}, which includes a selection of unpolarized and polarized beam measurements and recent fits. 

The resolution of this "form factor ratio puzzle" is crucial to advance our knowledge of the proton form factors, and with that, of the distribution of charge and magnetization inside the proton.

\section{TWO-PHOTON EXCHANGE}
As a possible explanation for the form factor ratio discrepancy, Blunden et al.\ \cite{Blunden:2003sp} suggested that hard two-photon exchange (TPE), neglected in standard radiative corrections, could be an important effect in Rosenbluth-type experiments. Two-photon exchange corresponds to a group of diagrams in the second order Born approximation of lepton scattering, namely those where two photon lines connect the lepton and proton. The so-called ``soft'' case, when one of the photons has negligible momentum, is included in the standard radiative corrections, like ref.\ \cite{MoTsai,Maximon2000}, to cancel infrared divergences from other diagrams. The ``hard'' part, where both photons can carry considerable momentum, is not. It is important to note here that the division between soft and hard part is arbitrary, and different calculations use different prescriptions. 

\subsection{Theoretical calculations}
A full description of the available theoretical calculations are outside of the scope of this paper. Suffice it to say that they can be roughly divided into two groups: hadronic calculations, e.g.\ \cite{Blunden:2017nby}, which should be valid for $Q^2$ from 0 up to a couple of GeV$^2$, and GPDs based calculations, e.g.\ \cite{Afanasev:2005mp}, which should be valid from a couple of GeV$^2$ and up. 

\subsection{Phenomenological extraction}
Assuming that TPE is indeed the only (or at least the dominant) source of the discrepancy, it is possible to extract the size of the effect directly from the available data. For example, in \cite{bernauer13}, the authors built a model based on the following assumptions:
\begin{itemize}
\item TPE dominantly affects the Rosenbluth-type experiments, leaving polarization data unchanged.
\item The effect is roughly linear in $\epsilon$. This is supported by the fact that no strong deviations from a straight line have been found in Rosenbluth separations so far.
\item The effect vanishes for forward scattering, i.e., for $\epsilon=1$.
\item For $Q^2\rightarrow 0$, TPE is given by the Feshbach Coulomb correction \cite{McKinley:1948zz}. Modern theoretical calculations have the same limit.
\end{itemize}
Assuming a correction of the form $1+\delta_{TPE}$ to the cross section, with 
\begin{equation}
\delta_{TPE}=\delta_\mathsf{Feshbach}+a(1-\epsilon)\ln{(1+b*Q^2)},\label{eqfesh}
\end{equation}
the authors could fit the combined world data set with excellent $\chi^2$. This extraction will be used in the following to predict the size of the effect.

\section{THE CURRENT STATUS}
Three contemporary experiments have tried to measure the size of TPE, based at VEPP-3 \cite{Rachek:2014fam}, Jefferson Lab (CLAS, \cite{Adikaram:2014ykv}) and DESY (OLYMPUS, \cite{Henderson:2016dea}) The next order correction to the first order Born calculation of the elastic lepton-proton cross section contains terms corresponding to the product of the diagrams of one-photon and two-photon exchange. These terms change sign with the lepton charge sign. It is therefore possible to determine the size of TPE by measuring the ratio of positron to electron scattering: $R_{2\gamma}=\frac{\sigma_{e^+}}{\sigma_{e^-}}\approx 1+2\delta_{TPE}$.

\begin{figure}[htb]
  \centerline{\includegraphics[width=0.5\textwidth]{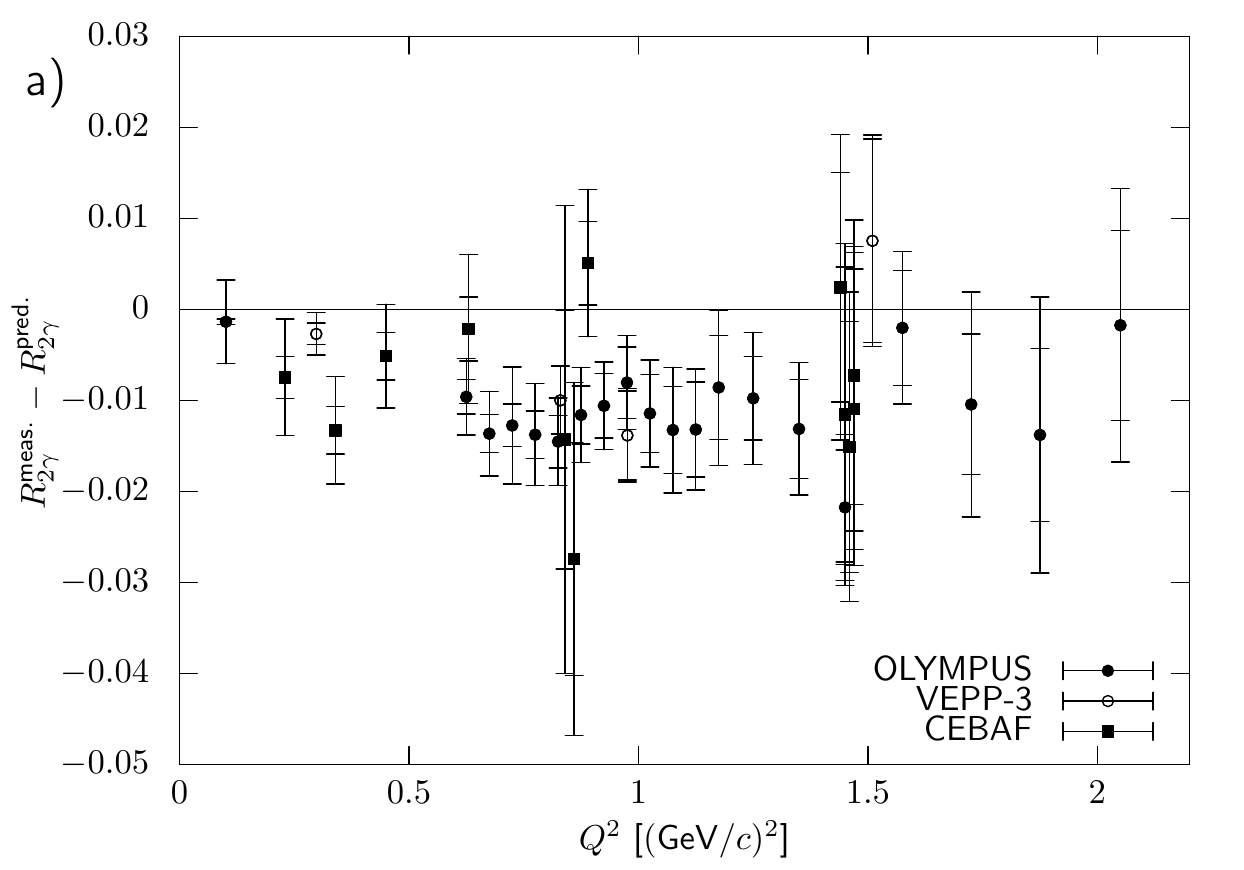}\includegraphics[width=0.5\textwidth]{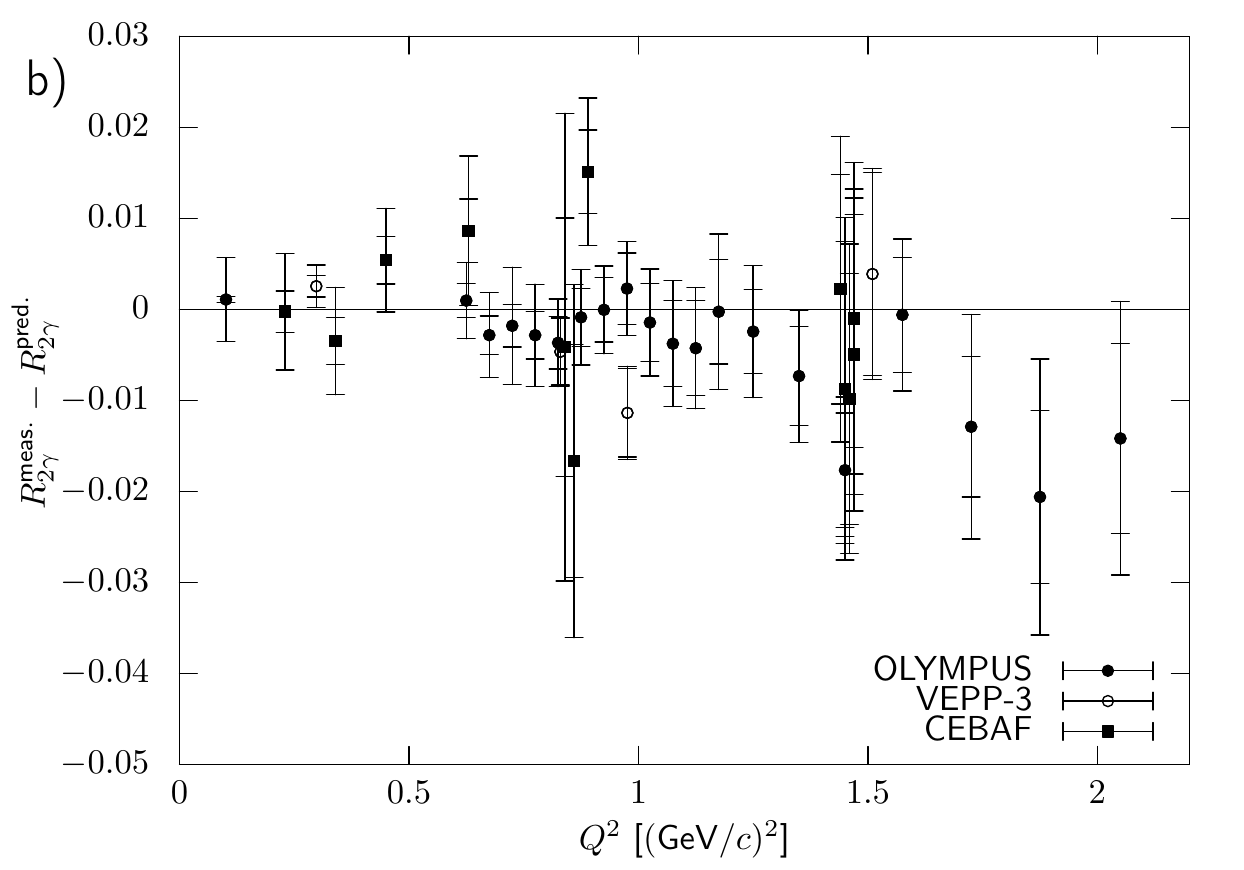}
  }
  \caption{\label{figdiff} Difference of the data of the three recent TPE experiments \cite{Rachek:2014fam,Adikaram:2014ykv,Henderson:2016dea} to the calculation in \cite{Blunden:2017nby} (a) and the phenomenological prediction from \cite{Bernauer:2013tpr} (b).}
\end{figure}

Figure \ref{figdiff} depicts the difference of the data of the three experiments to the calculation by Blunden et al.\ \cite{Blunden:2017nby} and the phenomenological prediction by Bernauer et al.\ \cite{Bernauer:2013tpr}. It can be seen that the three data sets are in good agreement which each other, and appear about 1\% low compared to the calculation. The prediction appears closer for most of the $Q^2$ range, however over-predicts the effect size at large $Q^2$. This is worrisome, as this coincides with the opening of the divergence in the fits depicted in Fig.\ \ref{figratio} and might point to an additional effect beyond TPE that drives the difference.

The combination of the experiments prefer the phenomenological prediction with a reduced $\chi^2$ of 0.68, the theoretical calculation achieves a red.\ $\chi^2$ of 1.09, but is ruled out by the normalization information of both the CLAS experiment and OLYMPUS to a 99.6\% confidence level. No hard TPE is ruled out with a significantly worse red.\ $\chi^2$ of 1.53.

The current status can be summarized as such:
\begin{itemize}
\item TPE exists, but is small in the covered region.
\item Hadronic theoretical calculations, supposed to be valid in this kinematical regime, might not be good enough yet.
\item Calculations based on GPDs, valid at higher $Q^2$, are so far not tested at all by experiment.
\item A comparison with the phenomenological extraction allows for the possibility that the discrepancy might not stem from TPE alone.
\end{itemize}

We refer to \cite{Afanasev:2017gsk} for a more in-depth review. The uncertainty in the resolution of the ratio puzzle jeopardizes the extraction of reliable form factor information, especially at high $Q^2$, as covered by the Jefferson Lab 12 GeV program. Clearly, new data are needed. In the following, we will discuss experimental possibilities.

\section{NEXT GENERATION EXPERIMENTS}
\subsection{Effect size and figure of merit}
As can be seen from Eq.\ \ref{eqfesh}, the size of TPE scales linearly with $1-\epsilon$, but only weakly with $Q^2$. The strongest signal is therefore at large $Q^2$ and small $\epsilon$. The cross section, however, drops fast in the same limit. To find the optimal kinematics, a figure of merit can be constructed by the ratio of expected deviation of $R_{2\gamma}$ from 1 and the expected uncertainty:
\begin{equation}
  FOM=\frac{\left|R_{2\gamma}-1\right|}{\sqrt{\Delta^2_\mathsf{stat.}+\Delta^2_\mathsf{syst.}}}
  \end{equation}
Here, we split the total uncertainty into a statistical and a systematical part. We fix the latter to 1\% for the following discussion.

Unfortunately, accelerators with positron beams in the relevant energy range are not very common. We discuss now the possibilities at two locations, DESY and Jefferson Lab.

\subsection{Measurement at DESY}
\begin{figure}[htb]
  \centerline{\includegraphics[width=0.5\textwidth]{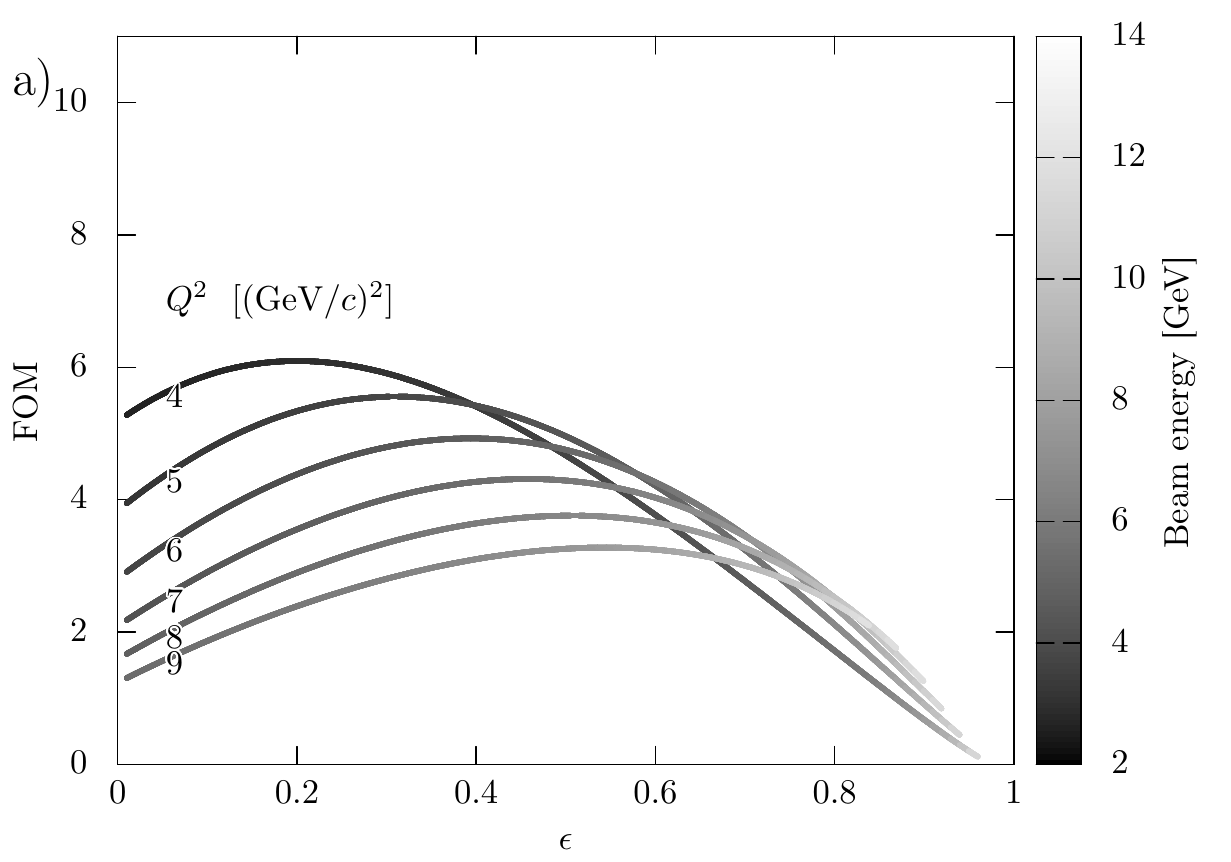}\includegraphics[width=0.5\textwidth]{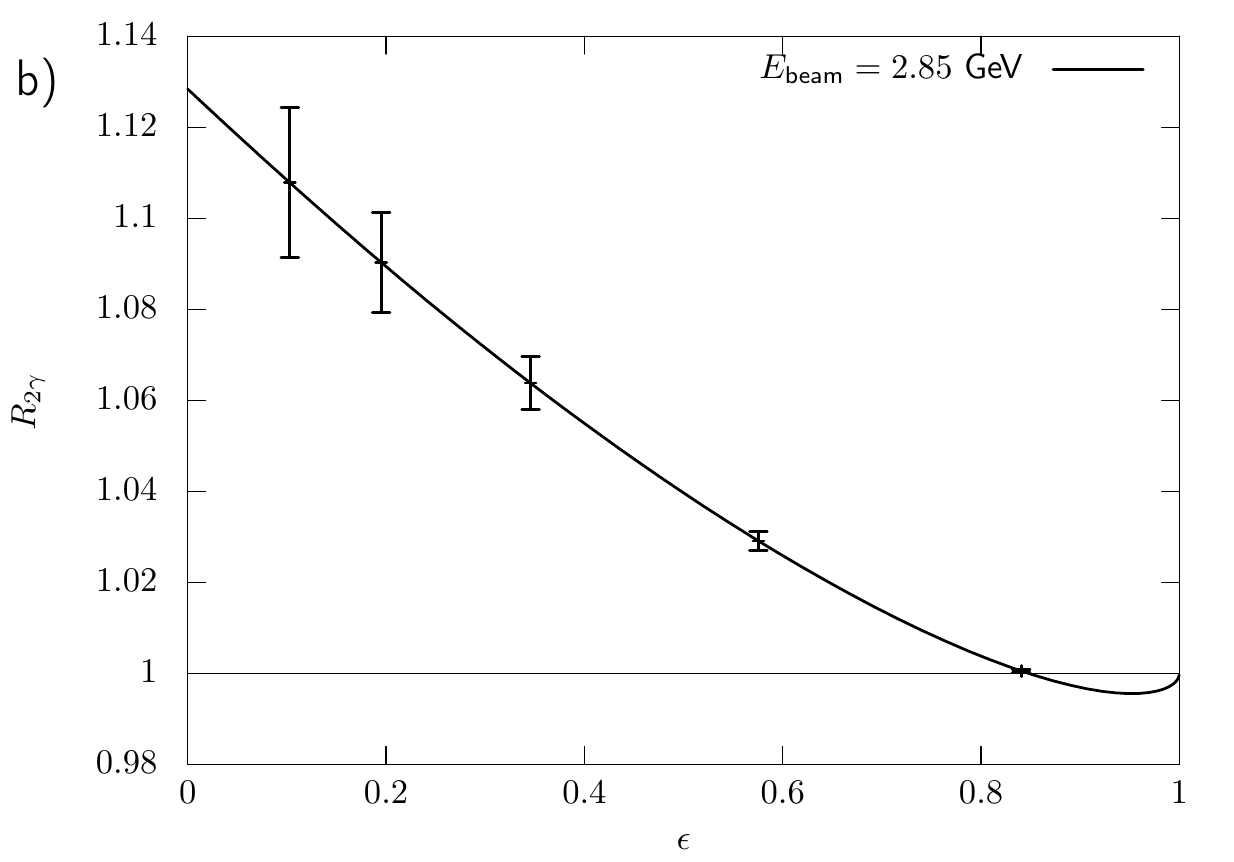}
  }
  \caption{\label{figdesy}a) Figure of merit as a function of $\epsilon$, for various $Q^2$, for 15 days of beam per species at DESY. b) Expected statistical error of data points and predicted effect size. }
\end{figure}

At DESY, a test beam facility is currently being proposed which would allow to do TPE measurements with a \SI{60}{\nano\ampere} beam. Due to constraints of allotted space and installation time, non-magnetic calorimetric detectors, such as the those designed for PANDA, would make suitable detectors. We assume five detector elements covering 10 msr each. The beam impinges on a 10 cm liquid hydrogen target. Figure \ref{figdesy} a) shows the FOM plot for 15 days per species. With a 2.85 GeV beam, the experiment could test TPE up to a $Q^2$ of about 4.5 GeV$^2$ with more than 5$\sigma$. The projected errors for such a measurement are shown in Fig.\ \ref{figdesy} b).

\subsection{Measurement at Jefferson Laboratory}
Jefferson Lab is evaluating the construction of a positron source for CEBAF. We assume that such a source would enable CEBAF to deliver up to \SI{1}{\micro\ampere} of unpolarized positrons into the experiment halls. We further assume a 10 cm liquid hydrogen target. This combination yields a luminosity of $\mathcal{L}=\SI{2.6}{\per\pico\barn\per\second}$.

\begin{figure}[htb]
  \centerline{\includegraphics[width=0.5\textwidth]{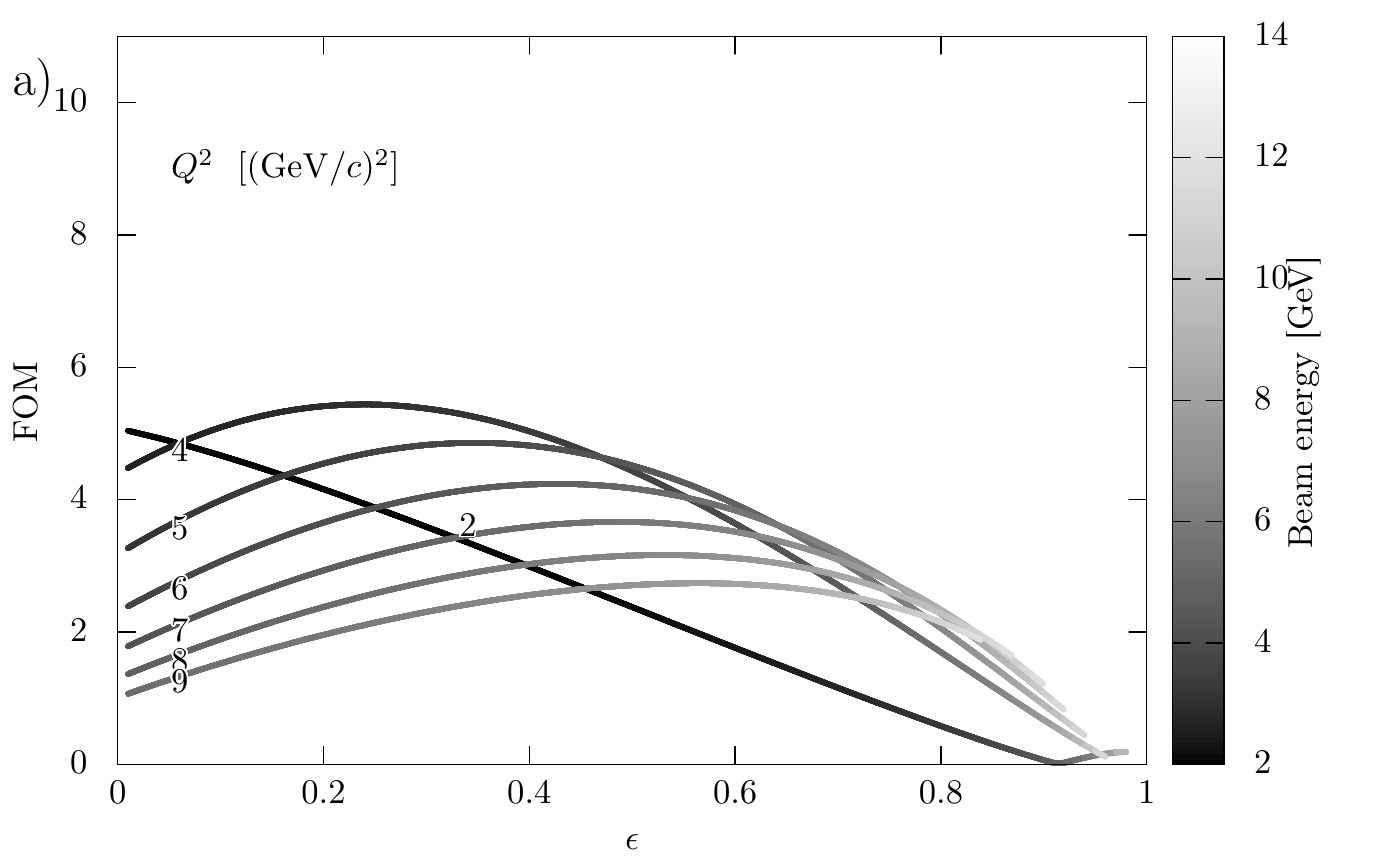}\includegraphics[width=0.5\textwidth]{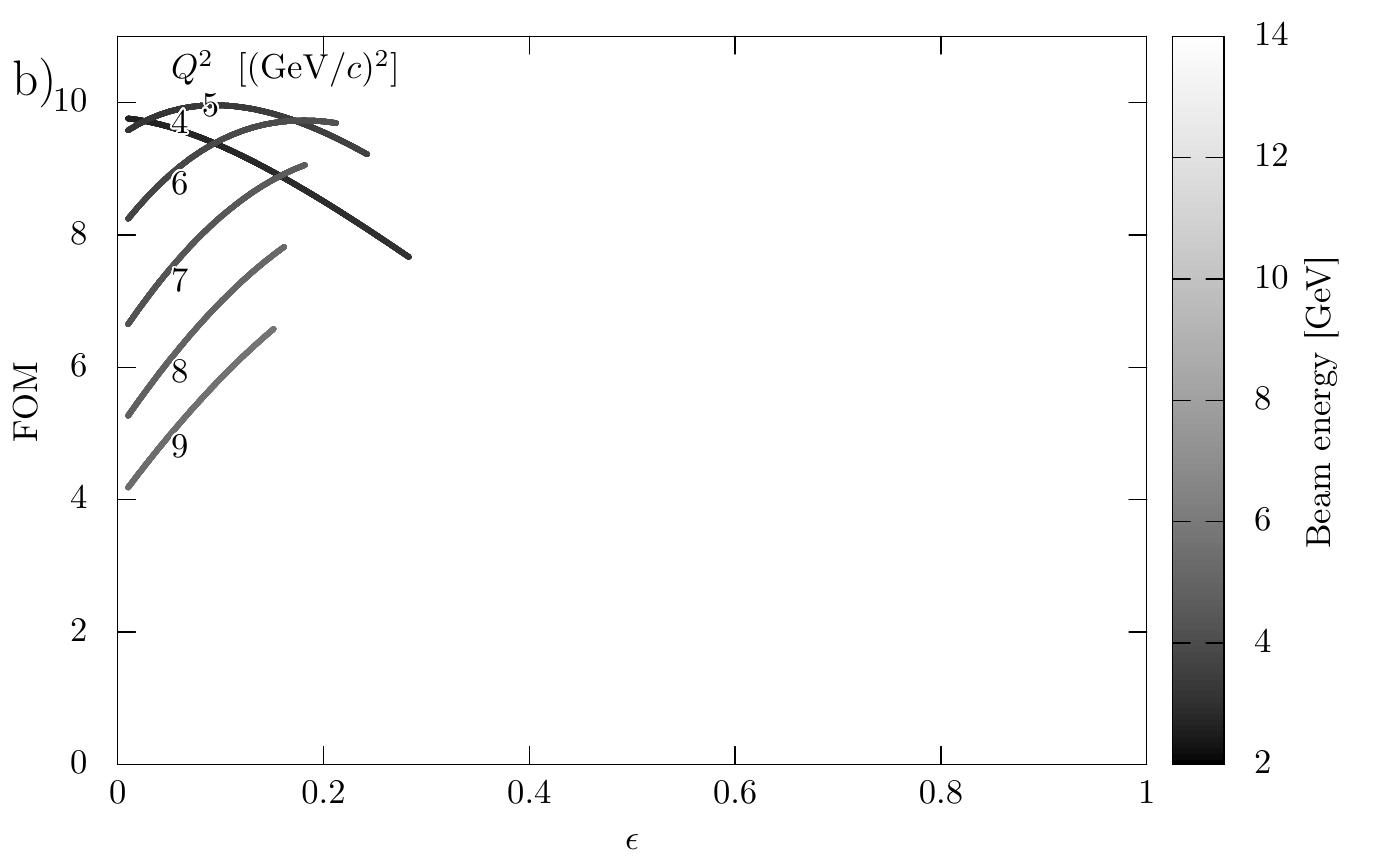}
  }
  \caption{\label{figfomjlab}Figures of merit as a function of $\epsilon$, for various $Q^2$, for 1 day of beam per species at Jefferson lab. Small acceptance spectrometers are represented by a), BigBite is represented by b).}
\end{figure}

For the purpose of this work, we looked at the possibilities in Hall A and C. The main spectrometers of Hall A, with 6.7 msr acceptance, and the HMS spectrometer in Hall C  are very versatile. The SHMS in Hall C is limited to forward angles, but could be used to detect the protons instead of the leptons, with the benefit of different systematical uncertainties. BigBite in Hall A is limited in the maximum momentum. However, the large acceptance allows measurements at very low values of $\epsilon$ with excellent figures of merit. Figure \ref{figfomjlab} depicts the figure of merit for 1 day per species, with the smaller-acceptance spectrometers represented by sub-figure a), and BigBite by sub-figure b).

A sketch of a possible measurement program for Hall A and Hall C is listed in Tab.\ \ref{tabhalla} and Tab.\ \ref{tabhallb}, respectively.

 \begin{table}[htb]
\caption{Proposed measurement program for Hall A}
\label{tabhalla}
\tabcolsep7pt\begin{tabular}{lcccccccccc}
\hline
  \tch{1}{c}{b}{$E_\mathrm{beam}$}  & \tch{3}{c}{b}{\SI{3.1}{\giga\electronvolt}}  & \tch{3}{c}{b}{\SI{3.55}{\giga\electronvolt}}  & \tch{3}{c}{b}{\SI{4.01}{\giga\electronvolt}}   \\
\hline
Spectrometer angles\tabnoteref{t1n1} & $30^\circ$ & $70^\circ$ & $110^\circ$ & $52.7^\circ$ & $70^\circ$ & $110^\circ$ & $42.55^\circ$ & $70^\circ$ & $110^\circ$\\
$Q^2$ [$(\mathrm{GeV}/c)^2)$]   & 1.79 & 3.99 & 4.75 & 3.99 & 4.75 & 5.56 & 3.99 & 5.55 & 6.4\\
$\epsilon$ & 0.82 & 0.32 & 0.1 & 0.49 & 0.30 & 0.09 & 0.60 & 0.28 & 0.08\\
Time [days/species] & \multicolumn{3}{c}{1}& \multicolumn{3}{c}{2}& \multicolumn{3}{c}{3}\\
\hline
\end{tabular}
\tablenote[t1n1]{Central angles of the two main spectrometers followed by the central angle of BigBite.}
 \end{table}

  \begin{table}[htb]
\caption{Proposed measurement program for Hall B}
\label{tabhallb}
\tabcolsep7pt\begin{tabular}{lccccccc}
\hline
  \tch{1}{c}{b}{$E_\mathrm{beam}$}  & \tch{2}{c}{b}{\SI{3.1}{\giga\electronvolt}}  & \tch{2}{c}{b}{\SI{3.55}{\giga\electronvolt}}  & \tch{2}{c}{b}{\SI{4.01}{\giga\electronvolt}}   \\
\hline
Spectrometer angles\tabnoteref{t2n1} & $79.7^\circ$ & $7.64^\circ$ ($120^\circ$) & $70^\circ$ & $9.95^\circ$ ($100^\circ$) & $18^\circ$ & $16.57^\circ$ ($65^\circ$) \\
$Q^2$ [$(\mathrm{GeV}/c)^2)$]   & 4.25 & 4.84 & 4.76 & 5.43 & 1.3& 5.35\\
$\epsilon$ & 0.244& 0.06 & 0.302& 0.122 & 0.935 & 0.33\\
Time [days/species] & \multicolumn{2}{c}{3}& \multicolumn{2}{c}{2}& \multicolumn{2}{c}{1}\\
\hline
\end{tabular}
\tablenote[t2n1]{Central angles for HMS (leptons) and SHMS (protons), with the equivalent lepton angle in parenthesis.}
\end{table}

\begin{figure}[htb]
  \centerline{\includegraphics[width=0.5\textwidth]{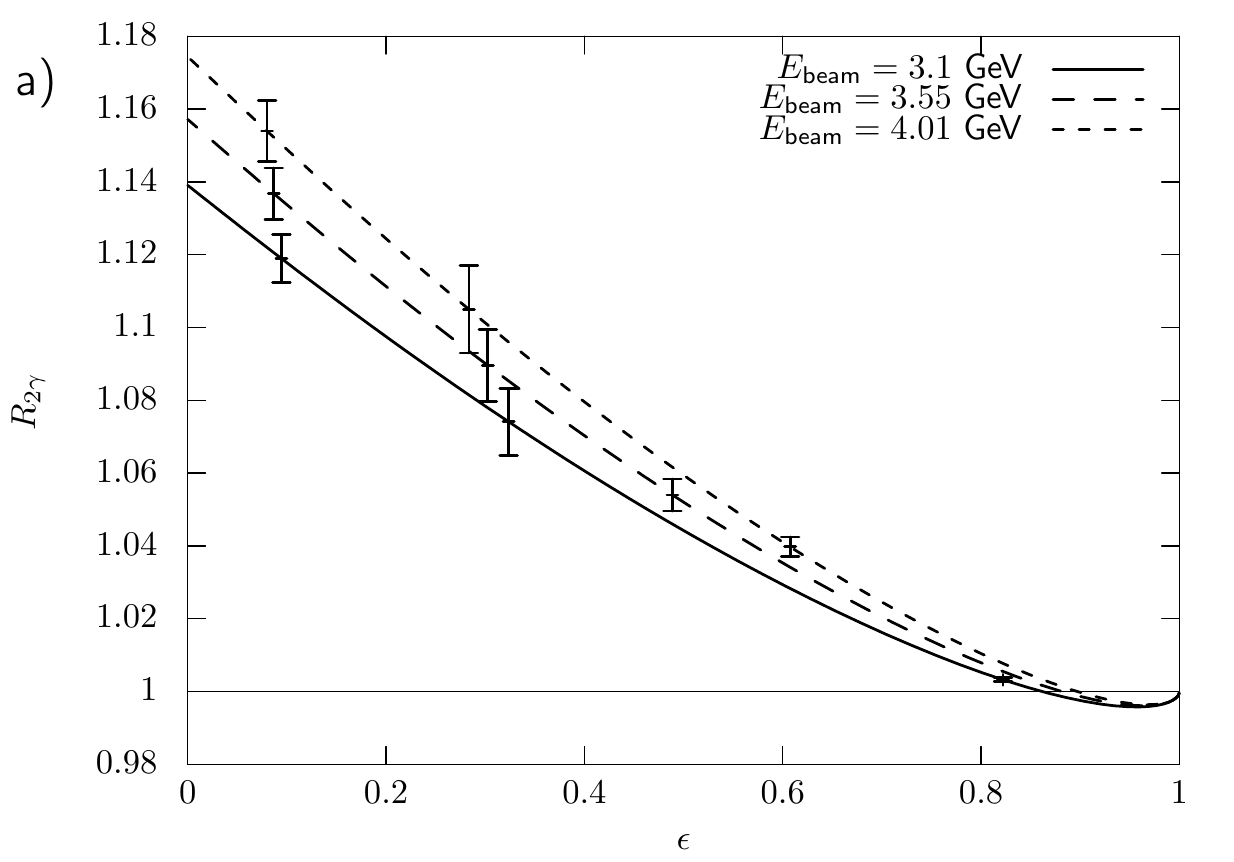}\includegraphics[width=0.5\textwidth]{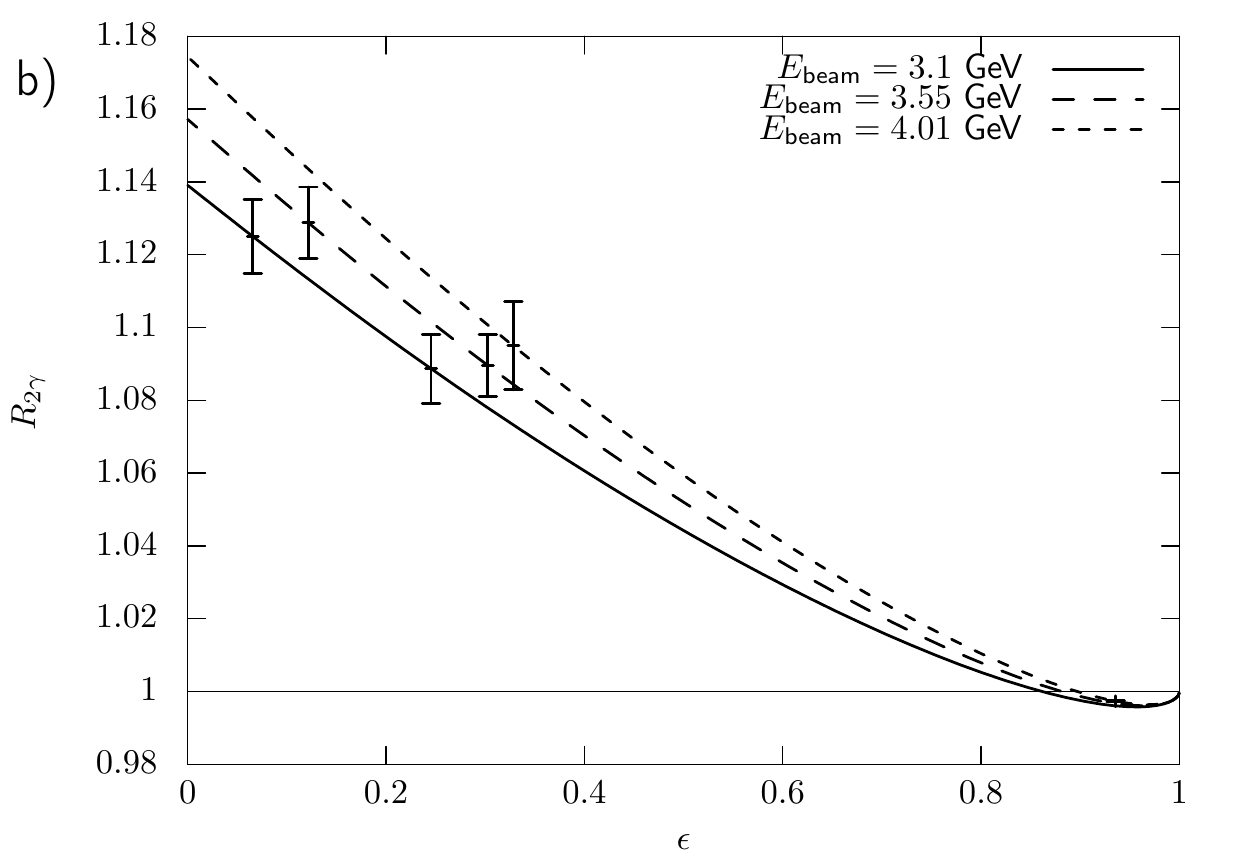}
  }
  \caption{\label{figerrjlab}Predicted effect size and estimated errors for the proposed measurement program in Hall A (a) and Hall C (b).}
\end{figure}

Figure \ref{figerrjlab} show the estimated errors and predicted effect size for Hall A (a) and Hall C (b). A high-impact measurement is possible with a comparatively small amount of beam time. Even in the case the final positron beam current is lower than assumed here, the experiment remains feasible.

\subsection{Systematic errors}
The main benefit to measure both lepton species in the same beam time is the cancellation of many systematics which would affect the result if data of a new positron scattering measurement is compared to existing electron scattering data. For example, one can put tighter limits on the change of detector efficiency and acceptance changes between the two measurements if they are close together in time, or optimally, interleaved. 

For the ratio, only relative effects between the species types are relevant; the absolute luminosity, detector efficiency, etc.\ cancel. Of special concern here is the luminosity. While an absolute luminosity is not needed, a precise determination of the species-relative luminosity is crucial.  Precise relative measurement methods, for example based on M\o ller scattering, exist, but only work when the species is not changed. Switching to Bhabha scattering for the positron case and comparing with M\o ller scattering is essentially as challenging as as an absolute measurement. More suitable is a measurement of the lepton-proton cross section itself at extreme forward angles, i.e., $\epsilon\approx1$, where TPE should be negligible small and the cross section is the same for both species. In OLYMPUS, a method exploiting the detection of multiple scattered particles from the same beam bunch has been used with great success \cite{Schmidt:2017jby}.

\section{CONCLUSION}
The discrepancy in the form factor ratio is a serious obstacle in the exact determination of the proton form factors and a dedicated measurement program is needed to address this pressing issue. The proposed test beam area at DESY could host a first experiment to investigate TPE at larger momentum transfers on a short time line, but is ultimately limited in reach by the luminosity. At Jefferson Lab, an upgraded CEBAF would make more precise experiments at even larger momentum transfers possible. This would test both hadronic and GPD-based theoretical calculations of TPE, and allow us to extract a phenomenological model precise enough to analyze contemporary and future form factor measurements.


\section{ACKNOWLEDGMENTS}
This work was supported by the Office of Nuclear Physics
of the U.S. Department of Energy, grant No. DE-FG02-94ER40818.


\nocite{*}
\bibliographystyle{aipnum-cp}%
\bibliography{jpos}%

\end{document}